\documentclass{article}

\usepackage[final]{neurips_2021_ml4ps}

\usepackage{natbib}
\setcitestyle{square,comma}
\setcitestyle{numbers}

\usepackage{graphicx} % Required for including images

\usepackage[utf8]{inputenc} % allow utf-8 input
\usepackage[T1]{fontenc}    % use 8-bit T1 fonts
\usepackage{hyperref}       % hyperlinks
\usepackage{url}            % simple URL typesetting
\usepackage{booktabs}       % professional-quality tables
\usepackage{amsfonts}       % blackboard math symbols
\usepackage{nicefrac}       % compact symbols for 1/2, etc.
\usepackage{microtype}      % microtypography
\usepackage{xcolor}         % colors
\usepackage{amsmath}
\usepackage{amssymb}
\usepackage{amsfonts}

\title{Unsupervised Spectral Unmixing For Telluric Correction Using A Neural Network Autoencoder}

\author{%
  Rune D. Kjærsgaard \\
  DTU Compute\\
  \texttt{rdokj@dtu.dk} \\
   \And
   Aaron Bello-Arufe \\
   DTU Space\\
   \texttt{aarb@space.dtu.dk} \\
   \And
   Alexander D. Rathcke \\
   DTU Space\\
   \texttt{rathcke@space.dtu.dk} \\
   \AND
   Lars A. Buchhave \\
   DTU Compute\\
   \texttt{buchhave@space.dtu.dk} \\
   \And
   Line K. H. Clemmensen \\
   DTU Compute\\
   \texttt{lkhc@dtu.dk} \\
}

\begin{document}

\maketitle

\begin{abstract}
The absorption of light by molecules in the atmosphere of Earth is a complication for ground-based observations of astrophysical objects. Comprehensive information on various molecular species is required to correct for this so called telluric absorption. We present a neural network autoencoder approach for extracting a telluric transmission spectrum from a large set of high-precision observed solar spectra from the HARPS-N radial velocity spectrograph. We accomplish this by reducing the data into a compressed representation, which allows us to unveil the underlying solar spectrum and simultaneously uncover the different modes of variation in the observed spectra relating to the absorption of $\mathrm{H_2O}$ and $\mathrm{O_2}$ in the atmosphere of Earth. We demonstrate how the extracted components can be used to remove $\mathrm{H_2O}$ and $\mathrm{O_2}$ tellurics in a validation observation with similar accuracy and at less computational expense than a synthetic approach with \texttt{molecfit}.
\end{abstract}

\section{Introduction}
\label{sec:intro}
Absorption of light in the atmosphere of Earth, called telluric absorption, can hinder astrophysical observations by partially obscuring the object of interest. Various methods have been introduced to remove the effects of this absorption from observed spectra. One such acknowledged method, called \texttt{molecfit} \citep{Molecfit1,Molecfit2}, relies on computing a synthetic transmission spectrum of the atmosphere of Earth. Synthetic methods are inherently reliant on external factors to an observation, such as atmospheric measurements and molecular line lists. Another realm of methods take a data-driven approach attempting to exploit the modes of variation in a number of observed spectra to uncover the underlying components. By analysing such variation, the telluric absorption can be modelled without relying on external factors to an observation. One such approach, based on principal component analysis (PCA), has been explored in the literature \citep{PCA_telluric}. PCA methods are however ineffective on very large data sets, where the entire data can not be stored in memory. Another approach called wobble \citep{bedell2019wobble} uses a linear model with a convex objective to model the telluric component of observed spectra. 

We present a new data-driven approach using a neural network autoencoder. Autoencoders have seen use in the literature for decades \citep{AEhistory1,AEhistory2} and have long been known to discover effective compressed data representations through dimensionality reduction \citep{AEDimReduction}. To demonstrate the approach we analyse 1257 observed solar spectra \citep{HARPS-N-Download,HARPS_paper} from the high-precision spectrograph HARPS-N \citep{HARPS-N} with the aim of disentangling the observed spectra into an underlying solar component and high accuracy telluric components from $\mathrm{H_2O}$ and $\mathrm{O_2}$. The extracted components could aid in the detection of radial velocity signals of planetary systems by quickly and accurately removing tellurics from observed spectra, leading to an increase in observation quality and hereby a reduction in observing time and cost.

\section{Proposed approach}
\label{sec:Network}
Our approach has roots in spectral unmixing, which seeks to unmix distinct endmember spectra and their weights from an observed spectral image by constructing a mixing model of the problem. Endmember unmixing from spectral data is a rich discipline with many existing approaches \citep{somers2011endmember,Review-Non-Linear-HSU}. We consider the solar spectrum and telluric spectrum as endmember spectra with associated abundance weights and use these components to construct a linear mixing model in log-space \citep{log_linear} describing the observed spectra:

\begin{equation}
\label{Eq. LMM Matrix format}
     \boldsymbol{x}_n = \sum_{r=1}^R  w_{r,n} \boldsymbol{m}_r = \boldsymbol{Mw}_n+\boldsymbol{\epsilon}_n,
\end{equation}
where $\boldsymbol{x}_n$ is the $n^{th}$ observed spectrum from a finite set of $N$ observed spectra, $\boldsymbol{m}_r$ is the $r^{th}$ endmember spectrum of $R$ endmembers with individual endmembers $r = 1,...,R$. Furthermore, $w_{r,n}$ is the abundance of endmember $r$ for observation $n$, $\boldsymbol{M}$ is the endmember matrix having endmembers as columns, $\boldsymbol{w}_n$ is the abundance vector of the $n^{th}$ observation and $\boldsymbol{\epsilon}_n$ is an error term. The HARPS-N spectra cover the optical wavelength range and for this reason we consider the combined telluric spectrum to be comprised of the two strongest absorbing molecules in this region, namely $\mathrm{H_2O}$ and $\mathrm{O_2}$. From this we get $R=3$ with $r=1$ representing the solar endmember, $r=2$ representing the $\mathrm{H_2O}$ endmember and $r=3$ representing the $\mathrm{O_2}$ endmember. 

\begin{figure}[h!]
  \centering
  \includegraphics[width=0.80\textwidth]{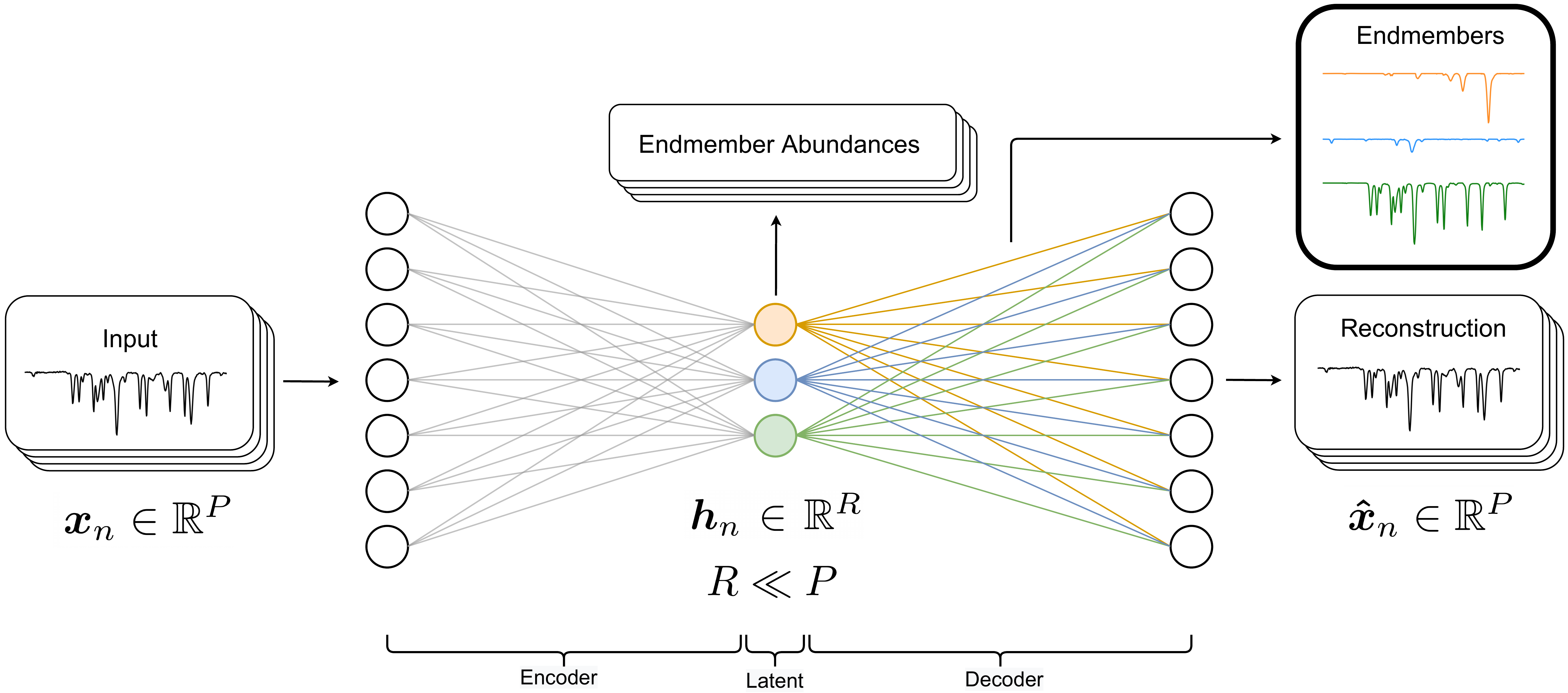}
  \caption{The figure shows the architecture of the autoencoder. Observed spectra $\boldsymbol{x}_n$ are given as input and passed through the encoder into a lower dimensional latent space, which is subsequently decoded into the reconstruction $\hat{\boldsymbol{x}}_n$. After training by minimising the reconstruction error through a gradient descent algorithm, the endmember matrix $\boldsymbol{M}$ is extracted as the weights of the decoder and the abundance vector $\boldsymbol{w}_n$ is extracted as the latent representation $\boldsymbol{h}_n$. $P$ is the number of pixels for each aperture in the observed spectrum. The network is illustrated for $R=3$ endmembers representing the solar (orange, top), $\mathrm{H_2O}$ (blue, middle) and $\mathrm{O_2}$ (green, bottom) endmembers.}
  \label{fig:AE}
\end{figure}

The goal is to extract the endmember matrix $\boldsymbol{M}$ and abundance vector $\boldsymbol{w}_n$ by training the neural network on a set of preprocessed observed solar spectra, with the purpose of applying the extracted telluric spectrum to non-solar observations. The endmember matrix $\boldsymbol{M}$ is extracted for training regions with $P$ pixels. Figure \ref{fig:AE} shows a graphical representation of the approach. 

In \citep{HS_unmixing} they present an autoencoder for blind unmixing of hyperspectral images (HSI). We build on this idea by adapting the network architecture to the domain of astrophysical spectral data. This requires various structural changes and the introduction of several new constraints on the network.

The training data consists of 1257 observations of the solar spectrum all split into 69 apertures with $P = 4096$ pixels each. This gives a total of 69 $\times$ 4096 pixels per observation. We use solar data for training since this data has a very high signal to noise ratio and does not take away observing time from night time observations. The observations are from 2020 spanning a month of observations from October 22 through November 19. We filter out low flux and high airmass observations. Subsequently we interpolate all observations to a common wavelength grid, apply the natural logarithm and continuum normalise the spectra. The described procedures leave $N= 838$ spectra of which $50\%$ are reserved for training and $50\%$ are used for validation. 

\subsection{Neural network autoencoder}
\label{subsec:Neural Network Autoencoder}
The network uses fully connected layers and consists of an encoder function $\boldsymbol{h}_n=f(\boldsymbol{x}_n)$, which maps the input data $\boldsymbol{x}_n \in \mathbb{R}^P$ to an internal latent representation $\boldsymbol{h}_n \in \mathbb{R}^R$. This representation is then passed through the decoder function $g(\boldsymbol{h}_n)=\hat{\boldsymbol{x}}_n$, which seeks to reconstruct the input data $\boldsymbol{x}_n$ with the reconstruction $\boldsymbol{\hat{x}}_n \in \mathbb{R}^P$. The decoder is constructed without bias terms and performs the following affine transformation:
\begin{equation}
\label{Eq. Output Layer}
      \boldsymbol{\hat{x}}_n= \boldsymbol{W}\boldsymbol{h}_n,
\end{equation}

where $\boldsymbol{W} \in \mathbb{R}^{P \times R}$ are the weights of the decoder, which are extracted as the endmember spectra $\boldsymbol{M}$, and $\boldsymbol{h}_n \in \mathbb{R}^R$ is the latent representation, which can be extracted as the endmember abundances $\boldsymbol{w}_n$.

The network features a utility layer responsible for normalisation of endmember abundance variation, which ensures a fixed abundance of the solar component, and a utility layer clamping the solar decoder weights to interval $[0,1]$ and the telluric decoder weights to interval $[-1,0]$. Moreover, the network contains a utility layer Doppler shifting the solar decoder weights to account for the spectral shift caused by the rotation and elliptical orbit of the Earth. We perform this shift using the barycentric Earth radial velocity of each observation. The network also features a batch normalisation layer \citep{Bacthnorm1,Batchnorm2}. We use the validation set to determine hyperparameters based on a Tree-structured Parzen Estimator approach carried out with \texttt{optuna} \citep{optuna}. Training is performed with stochastic gradient descent to minimise the mean squared error (MSE). The network is trained on non-stitched spectra for the 69 apertures separately to retain the high fidelity of the observed spectra and to avoid the complications involved in stitching spectra. Training on all apertures takes approximately 3 hours and 15 minutes on an Intel 6 core i7, UHD 630 CPU laptop.

\section{Results}
\label{sec:results}
\begin{figure}[h!]
  \centering
  \includegraphics[width=0.75\textwidth]{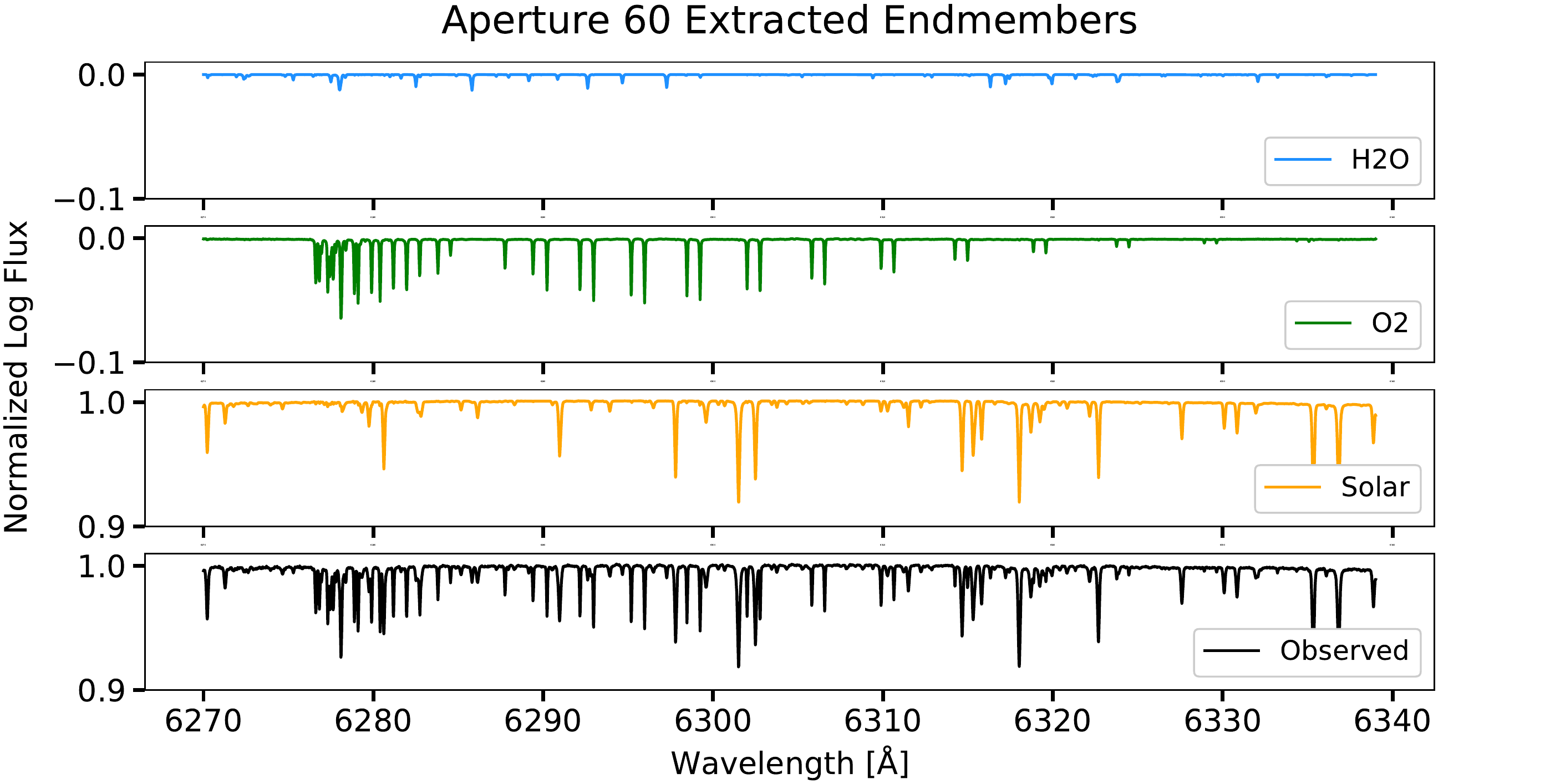}
  \caption{The figure illustrates extracted endmembers in addition to an observed solar spectrum for aperture 60. The extracted endmembers represent from top to bottom the $\mathrm{H_2O}$ (blue), $\mathrm{O_2}$ (green) and solar (orange) components of the observed spectrum (black, bottom).}
\label{fig:Order60}
\end{figure}

The extracted endmembers for aperture 60 are shown in Figure \ref{fig:Order60}.

\begin{figure}[h!]
  \centering
  \includegraphics[width=0.8\textwidth]{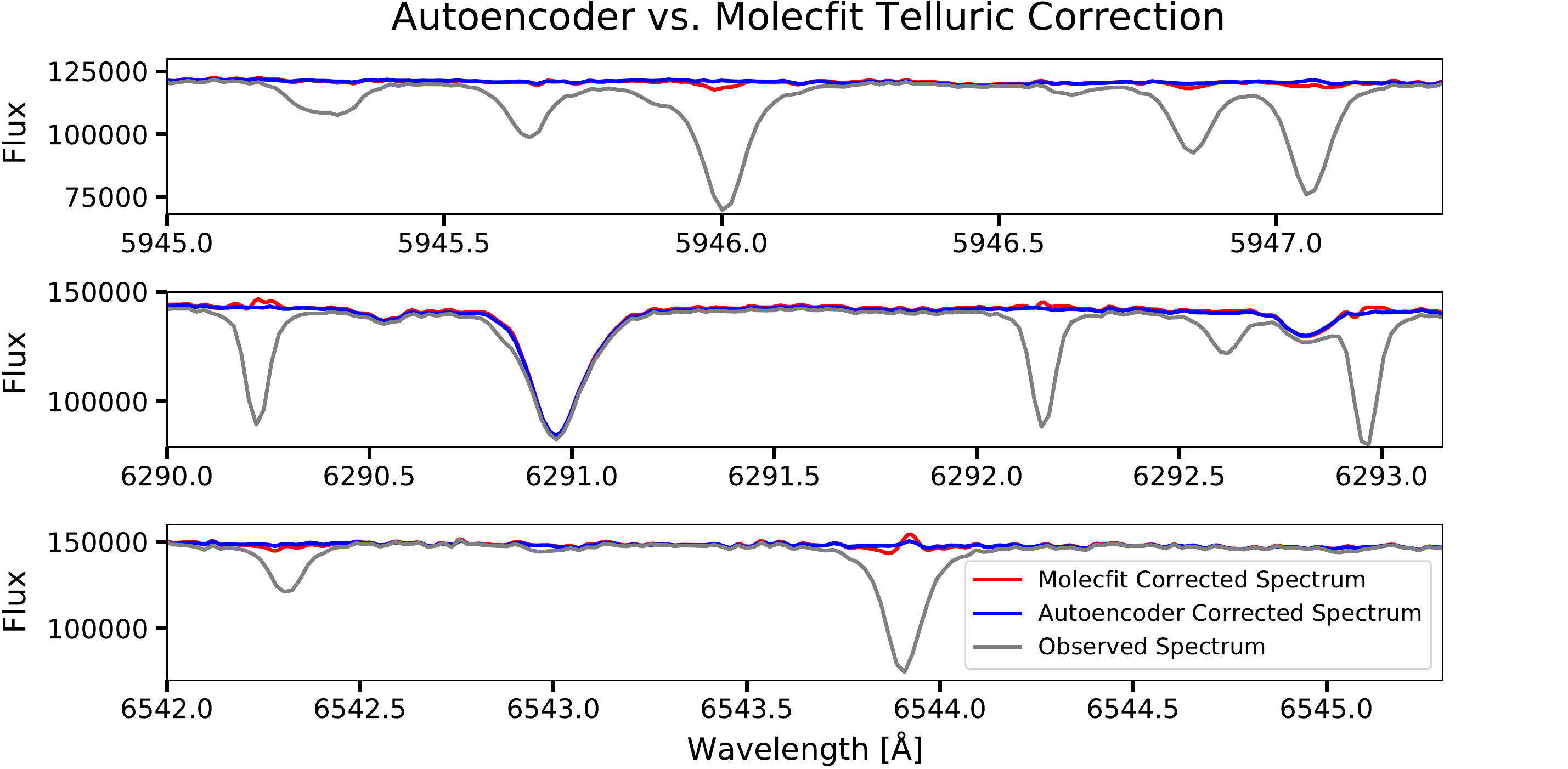}
  \caption{The figure displays a comparison of corrected spectra computed either using autoencoder extracted tellurics or \texttt{molecfit} tellurics for $\mathrm{H_2O}$ and $\mathrm{O_2}$ lines in three spectral regions of high interest to the study of exoplanetary atmospheres. For pure telluric lines, an ideal correction would result in a flat line. In the middle plot, the autoencoder and \texttt{molecfit} agree on the presence of solar features around 6290.5 Å, 6291 Å and 6292.75 Å, which remain in the corrected spectra.}
\label{fig:Correction}
\end{figure}

To validate extracted endmembers we perform telluric correction using our autoencoder tellurics and compare with a \texttt{molecfit} synthetic telluric spectrum. Telluric correction aims to remove tellurics by dividing the observed spectrum with a telluric transmission spectrum. We perform the comparison by correcting a solar observation with strong tellurics from HARPS-N, which the autoencoder has not been trained on. We compute the \texttt{molecfit} telluric spectrum on a HPC cluster using a 10 core Intel Xeon E5-2660v3, Huawei XH620 V3 node and utilise atmospheric measurements from the time of the observation in addition to a fit to the stitched version of the observation from the HARPS-N pipeline. We compute the autoencoder correction using the extracted telluric components, which have been converted back from log-space to represent standard transmission spectra. Autoencoder telluric abundance weights are found using a least squares fit to known telluric lines in the spectrum. We interpolate the observed spectrum to the telluric wavelength axes of \texttt{molecfit} and the autoencoder before the corrections. The comparison can be seen in Figure \ref{fig:Correction}.

\section{Discussion}
\label{sec:discussion}
High-resolution spectroscopy is limited to ground-based observations, and consequently high-resolution ground truth solar spectra are not available. This makes evaluating the performance difference between the two approaches complicated. A natural area of comparison however lies around corrected tellurics, where an accurate correction will leave no trace of the telluric absorption. As illustrated in Figure \ref{fig:Correction}, the autoencoder correction removes tellurics in the observed spectrum to continuum level, while \texttt{molecfit} leaves slight traces of the correction. The sinusoidal shape of the \texttt{molecfit} correction around 6544 Å in Figure \ref{fig:Correction} could indicate imprecision on the exact wavelength location of the telluric line centre used in \texttt{molecfit}. The \texttt{molecfit} correction takes approximately 30 minutes to compute, while the autoencoder correction takes about 0.2 seconds. This difference in compute time is significant and makes the autoencoder much more feasible for correction of multiple spectra. Additionally, the autoencoder approach can be used as a complementary data-driven validation tool to inspect the accuracy of synthetic approaches like \texttt{molecfit}.

While the results show that the autoencoder correction of the validation observation is performed with similar accuracy to \texttt{molecfit}, the question of how well the extracted endmembers generalise to observations dissimilar to the training data still remains. Future work includes exploring this potential limitation by performing corrections on numerous non-solar observations and inspecting the impact on radial velocity extraction, as well as the consistency of retrieved exoplanetary atmosphere signals.

This paper demonstrates the approach applied to the HARPS-N spectrograph, but the telluric autoencoder is designed as a general tool, which can be trained on solar data from any spectrograph and wavelength range and subsequently perform corrections on new observations (including non-solar observations) from the given spectrograph. This is an advantage as solar data is already gathered and ready to train on for many spectrographs. 

\section{Conclusion}
\label{sec:conclusion}
We have demonstrated an approach for computing a compressed representation of the solar data from HARPS-N with a constrained neural network autoencoder. This representation can be used to extract endmembers that directly relate to the solar spectrum as well as the transmission spectra of $\mathrm{H_2O}$ and $\mathrm{O_2}$. After the autoencoder representation has been computed for a given spectrograph, the extracted components can be used to perform very quick and accurate telluric correction on any observations from the same spectrograph. The autoencoder approach and the detailed extracted telluric spectrum could aid in the detection of faint radial velocity signals and atmospheric features of Earth analogue exoplanets observed from ground-based telescopes. 

\bibliography{refs}

\end{document}